# E-learning Services for Rural Communities

**Shariq Hussain**
School of Computer and Communication Engineering
University of Science and Technology Beijing
Beijing, China
s.hussain@zoho.com

**Zhaoshun Wang**
School of Computer and Communication Engineering
University of Science and Technology Beijing
Beijing, China
zhswang@sohu.com

**Sabit Rahim**
School of Automation and Electrical Engineering
University of Science and Technology Beijing
Beijing, China
sabit.rahim@kiu.edu.pk

## ABSTRACT
Information and communication technologies brought-in tools and techniques in the field of education that introduced new concepts of teaching and learning. Learning management system is one of the key tools used in educational institutes to facilitate e-learning. There is remarkable digital divide among urban and rural areas. In this paper, we present a model for providing e-learning services to remote/rural areas in order to promote and facilitate modern education. A dedicated resource center, hosting the learning management system, facilitates e-learning centers through Internet. The overall goal of this model is to have a cost-effective learning environment equipped with latest technologies to provide learners an opportunity to get insight into new information and communication technologies and e-learning environment. The model offers new teaching methodology with enhance utilization of learning management system in teaching and learning. Basic characteristics and technical aspects will be considered as well. The study will also promote development and usage of open-source technologies.

## 1. INTRODUCTION
The tremendous development in Information and Communication Technologies (ICTs) has paved the way for e-learning. Use of computers in education sector can be traced back to the early 1980s when simple word processors were in use [1]. The Internet has revolutionized the computer and communications world like nothing before. This brings us great learning opportunities by having access to large amount of information with benefits in terms of time and cost savings.

The modern educational technology facilitates design, delivery and management of educational activities for learners. This could be face-to-face in a lecture hall, online, or combination of both. Imparting education in this way is termed as e-learning (electronic learning) i.e. learning through information and communication technologies [2]. E-learning facilitates distance learning and provides means to learners to access learning material any time and at any place.

A learning management system (LMS) is the software application that facilitates e-learning [3]. Various commercial as well as open-source LMSs are available today which are being used in educational institutions.

Traditional learning involves setting up infrastructure of school/college and hiring of faculty and staff. Students have to attend school/college in order to learn. In the rural areas, development of such institute and hiring of full-time faculty requires a lot of resources. Professionally trained educators mostly prefer to work and reside in urban areas. Consequently, the population of rural areas is deprived of quality educators and thus quality education. The current work proposes development of e-learning centers based on ICT to provide good quality education with up-to-date learning material in rural areas. The model employs latest educational technologies that will improve the education standard and will provide means to introduce educational technology to learners. The learners will have a chance to get familiar with latest technologies which will give them deep understanding and effectiveness in using it. The study will promote e-learning platforms and also the development and usage of open-source technologies.

This paper is organized as follows: Section 2 presents an overview of e-learning and LMS is described in Section 3. Section 4 describes model for development of e-learning centers. Section 5 presents discussions and related work and Section 6 concludes the paper.

## 2. E-LEARNING
E-Learning is to revolutionize the learning process by usage of ICT resources [4]. E-learning is generally used in distance learning, but it can also be used in conjunction with face-to-face learning. Learning models are described below.

### 2.1 Learning Models
There are three main learning models. A brief description of each is given below.

#### 2.1.1 Traditional learning
Students have to attend lectures in a classroom. There is a face-to-face interaction between teachers and students [5]. Use of multimedia presentations can enhance the learning experience of students.

#### 2.1.2 Distance learning
Teachers and students are at different places for all or most of the time. Students are provided with pre-recorded, packaged learning materials and interaction between students and teachers take place through some form of communication technology [6]. This model also requires special organizational and administrative arrangements in order to provide an effective learning environment.

*2.1.3 Blended learning*

It is the combination of traditional learning model with e-learning solutions [7]. For example, learners attend a face-to-face session at the beginning and at the end of a program, with learning activities occurring online in the middle.

## 3. LEARNING MANAGEMENT SYSTEM

LMS is the software used for managing e-learning with a delivery mechanism, providing access to resources, tracking and assessing the academic activities. LMS provides an environment in which learning contents are developed and organized by instructors. Interactions between instructors and students take place through communication tools. Most of LMSs are web-based in order to facilitate online access to learning content. Features offered by LMS include user management, content management, activity tracking, file storage, storing of grades, reports generation, and communication tools (chatting, forums, discussion groups, email, teleconferencing, and so on). A typical LMS with some features is shown in Fig. 1.

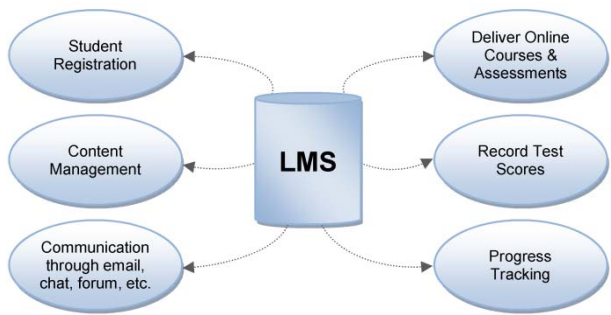

**Fig 1: Learning management system**

Instructors use authoring and publishing tools to create and publish learning contents. Courses are made more interesting and easy to understand with the help of enriched multimedia examples. The environment offer collaborative learning by assigning tasks to students, making assessments, assignment of grades, active participation with synchronous or asynchronous communication tools, such as chatting, discussion groups, forums, email, video conferencing, and so on. Access to learning resources is restricted to enrolled students. LMS is the core of proposed e-learning model as it will serve as platform for all learning activities.

## 4. MODEL FOR E-LEARNING SERVICES

In today's globalized world, Internet is the driving vehicle for content delivery systems [8]. Internet has eliminated the language and geographical barriers and facilitated in establishment and growth of international education community [9]. The tremendous development of ICT has influenced the education environment. It has brought educational technology in learning environments and provided new opportunities to get introduced to new horizons. The full benefits of e-learning can be explored by carefully designing and implementing e-learning environment. E-learning provides a collaborative learning environment to students and increases their engagement, motivation and helps them to become self-directed and independent learners [10]. Effective e-learning can help in spreading of modern education, improve the quality and equip learners with skills that lead them to contribute to better socio-economic development.

The proposed model is based on three entities, namely LMS resource center, e-learning center and learners.

### 4.1 LMS Resource Center

The central hub of the proposed model is LMS resource center. The resource center is established in city/town where ICT resources are available. This center is equipped with required resources to develop and distribute learning contents through Internet. The model of resource center is shown in Fig. 2. The main components of resource center are LMS server, LMS application, content development team and administrator. A brief description of each is given below.

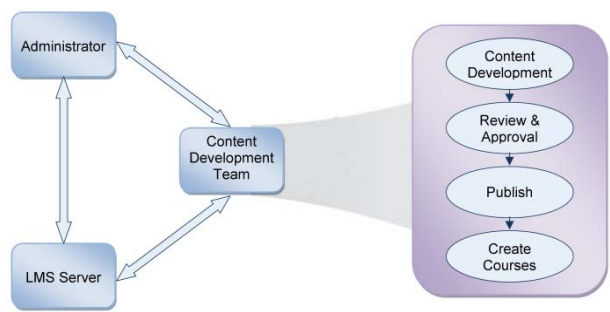

**Fig 2: LMS resource center**

*4.1.1 LMS server*

The LMS server consists of necessary hardware and software to process requests and delivers content to clients through Internet. LMS server is shown in Fig. 3. Freeware/Open-source software is selected for the proposed study, due to their free availability and zero cost.

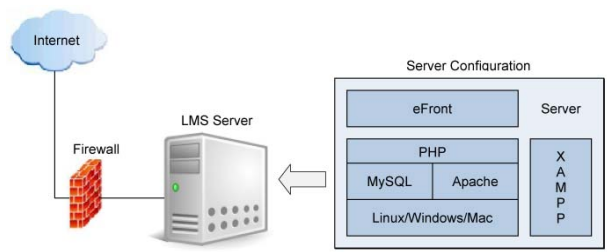

**Fig 3: LMS server**

Open-source software is not dependent on a single entity and gives users the advantage of the community. A large amount of developers globally contribute and analyze the code, making it more secure and constantly increasing the quality. Open-source provides flexibility, which is not available in closed products. There are a number of open-source LMS tools available, such as ATutor, Chamilo, Moodle, eFront, Claroline, Sakai and Bodington. The proposed e-learning model uses eFront LMS [11]. eFront LMS is discussed in later section. The hardware and software requirements that system must meet in order to install eFront LMS are presented in Table 1 and Table 2 respectively. It is difficult to install and configure (Apache/MySQL/PHP) AMP manually. The solution is XAMPP [12] that is an integrated server package of Apache [13], MySQL [14], PHP [15] and Perl [16]. XAMPP is freeware and configures the AMP environment in an automated way. Storage space requirements increase with the increase in number of users as each user requires storage space for files/data storage. The server can support fairly a large numbers of registered users and potentially 400–500 concurrent users (users online at any one point in time). Storage space and memory can be increased to add support for more users.

**Table 1. Hardware requirements for LMS server**

| Hardware | Requirements |
|---|---|
| Processor | Intel Duo Core Processor or higher |
| Memory (RAM) | 4 GB |
| Hard disk space | SAS/SATA RAID 1 (2 drives) 50GB, 10K RPM SCSI/SAS Hard Drive |

**Table 2. Software requirements for LMS server**

| Software | Requirements |
|---|---|
| Operating System | Linux/Windows/Mac |
| LMS | eFront Open-source version 3.6 |
| Web Server | Apache 2+ recommended, although eFront can work with Apache 1.x, IIS, nginx or lighttpd |
| Language | PHP version 5.1+ (PHP 5.2+ recommended) |
| Database | MySQL Server 4+ (MySQL 5 is strongly recommended) |

### 4.1.2 eFront LMS
Selection of eFront LMS is based upon the comparative study of open-source learning management systems [17]. From the study, it is observed that eFront has more visually attractive icon-based user interface that gives a nice look-and-feel and is easy to use. Most of the options are self-explanatory. eFront is capable of fulfilling a wide range of learning requirements by offering many tools for content management, quizzes, assignments, projects, reports, chat, forums, file sharing etc. Some of the features of eFront are shown in Fig. 4. eFront is user-friendly, extensible and suitable for both academic and organization use. eFront offers three user roles in its environment, namely Administrator, Professor and Student. User roles can be interchanged in different courses. eFront has multilingual support and can support more than forty languages.

### 4.1.3 Content development team
This team consists of Subject Leader, Instructor and Teaching Assistant (TA). The subject leader plans and prepares the course outline. After discussion with the instructor, guidelines are provided to TA for development of learning content. TA's role is to assist the subject leader and instructor in all academic activities. Course approval process is shown in Fig. 5. Once learning contents are prepared, the same are reviewed by subject leader and instructor. If there are some revisions required, assigned to TA otherwise forwarded to Principal for review and final approval. After finalization of course contents, the same are transformed into content package files which can be read and imported by LMS.

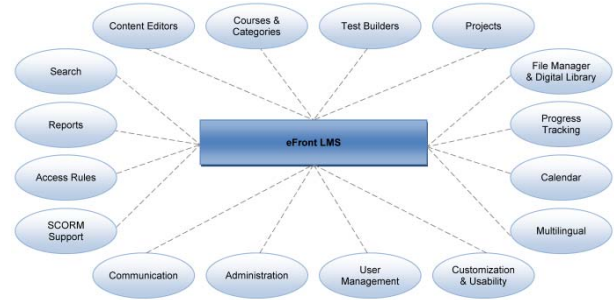

**Fig 4: Features of eFront LMS**

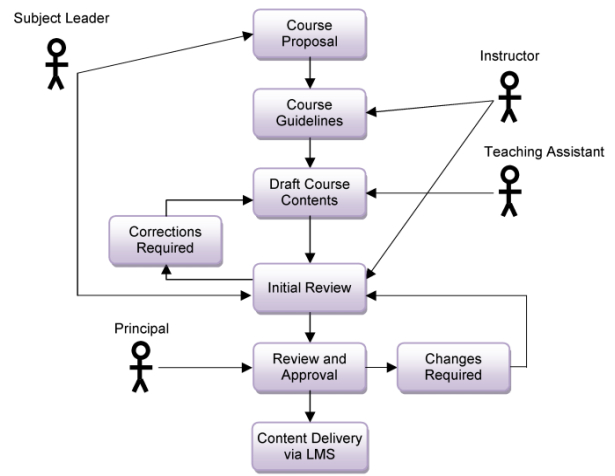

**Fig 5: Course approval process**

### 4.1.4 Administrator
The administrator can control any aspect of the system through an easy to use interface. The LMS Administrator works in liaison with instructor and reports to Principal. The LMS Administrator oversees operation and maintenance of LMS server, resolves user issues and assists with other help-desk duties. Due to the complexity of hardware and software that make up the LMS infrastructure, it is the responsibility of the LMS Administrator to perform any patches, upgrades, service packs, hot-fixes, and other routine maintenance to ensure the highest possible uptime and reliability of the LMS service.

## 4.2 E-learning Center
Target areas for establishing of e-learning centers are based upon the population rate and availability of communication infrastructure (in this case telephone lines). In the initial phase, schools can be used to setup e-learning centers. There is no activity taking place after school hours and further it will reduce the initial setup cost. Only requirement is to add IT equipment and communication media to setup e-learning facility for learners. A model of e-learning center is shown in Fig. 6.

The e-learning center is equipped with a computer having audio/visual accessories and connection to Internet through DSL/Cable modem. The e-learning center is supervised by an ICT Technician who is responsible for operation and maintenance of IT equipment in e-learning center. Initial training will be provided to ICT Technicians to have hands-on knowledge of LMS. The responsibilities of ICT Technician include:

- Provide technical support and resolve ICT related issues at e-learning center.
- Liaise with the Administrator at resource center regarding ICT matters.
- Ensure the continued operation of IT equipment at e-learning center.
- Perform software upgrades/patches after approval from Administrator.
- Keep up to date with ICT developments, and liaise with the Administrator over future developments, Internet and video conferencing.
- Provide assistance to instructor in delivering of lessons through LMS and guidance to learners in educational activities.

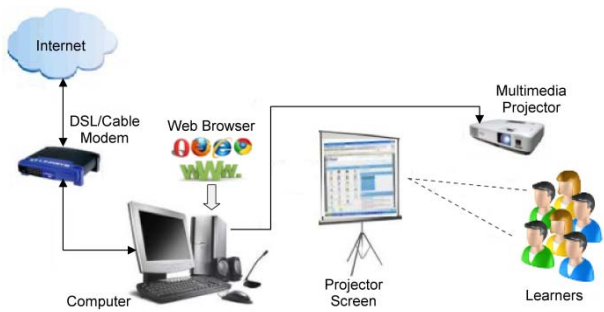

**Fig 6: E-learning center**

The basic goal of e-learning centers is to offer computer courses like basic IT essentials, programming, web development, network essentials, etc. In addition, language courses using e-learning environment is another interesting aspect. For example, English language [18], Chinese language [19] and many more language courses can be offered. Another aspect is use of e-learning for vocational training.

### 4.3 Learners
Learning is the process of acquiring or modifying knowledge, skills, and values by study [20]. Learner is the person who learns or takes up knowledge or beliefs. The learner uses the platform to get new competences. The target learners for e-learning centers are the students of age fifteen and above who have discontinued their study due to some reasons. Further, other students/persons who are interested in the courses to enhance their skills can also benefit from e-learning centers.

The proposed model is targeted for rural areas. In the developing countries, rural areas comprise of major population. Mostly these areas lack resources in access to health care, access to quality education, access to technology, transportation and communication. In these areas, it is observed that mostly students discontinue their study after secondary school. There are certain factors that have an effect on dropout ratio of students including low household income, family size, lack of good education environment, lack of guidance, non-availability of qualified human resource and resources.

It is also inferred that in rural areas children often leave study to become skilled workers so that they can contribute to their household income. They start doing low paying jobs, mostly laborers, helpers or attendants. The goal is to develop their interest in latest educational technology and equip them with proper skills. This may increase their ability to learn, experience and master latest technology and technological applications. By utilizing their basic education and learning skills they can find better jobs or start their own small business like computer sales, network design and maintenance, website development, software development etc. By doing so, they can contribute well in their income and raise their living standards and consequently contributing in the overall economy.

### 5. DISCUSSION AND RELATED WORK
The ICTs has a potentially huge role. The impact of ICTs in education has been assessed in various studies. The proper planning, design and implementation of online learning activities can lead to effective learning and contribution of ICTs to student performance. Online learning activities facilitate more effective education and offer significant benefits over traditional methods [21]. Studies revealed that use of ICTs in education have improved students' attitudes toward learning, development of teachers' technology skills and increased community's access to education and literacy [22]. Implementation of LMS in Greek high school showed that students developed better concepts, increased the amount of knowledge and use of their computers for knowledge acquisition [23]. The impact of socio-economic factors on girls and women's access to ICT education and training in a rural South African environment are studied and suggested strategies for improved access to ICT education and training [24]. The SPARK [25] project in Turkey is an example of ICT education for development. The goal of SPARK project is to improve the level of IT expertise among youth and it has shown significant results. ICTs are also being used in e-government projects. A study evaluated the effectiveness of ICT centers for developing e-governance in the rural communities in the west of Iran and proposed that ICT centers may also provide other services in order to become more effective in the process of rural development [26]. A comparative review of some case studies identified in literature is presented in Table 3. E-learning is being used within higher education institutes in different fields for conducting various courses in the form of blended learning model [18–19, 27–28]. E-learning can also be a valuable tool for special persons. People with disabilities can improve their education from home by use of e-learning [29].

In e-learning, the amount of interaction among the learners increases as compared to traditional learning environment. This is due to the fact that expressing oneself in computer mediated communication is more comfortable. More importantly, the level of interaction between the instructor and learners appears predominant in online learning.

There is a need to create more collaborative environment for learners for attainment of good results. As with all fundamental skills, the earlier the education system allows students to become familiar with technology the greater will be their depth of understanding and effectiveness in using it. In order to get the maximum benefits of ICTs, more financial resources for acquisition of resources and infrastructure for the promotion of e-learning facilities should be devoted.

Table 3. Comparative overview of different case studies

| Comparative Factors | Proposed Model | Models Proposed by Researchers in Literature | | | | | |
|---|---|---|---|---|---|---|---|
| | | [30] | [31] | [32] | [33] | [34] | [35] |
| Target Community | Rural | Remote Areas | Rural | Rural | Rural | Rural | Rural |
| Target Area | Vocational Training | Healthcare | Schools | Tourism | Business | Agriculture | Agriculture |
| Target Audience | Dropout Students | Children | Enrolled Students | Local People | Local Businessmen | Local Farmers | Local Farmers |
| Age Range | 15 and above | < 5 years | 4 to 18 years | Not specific | Not specific | Not specific | Not specific |
| Delivery System | Learning Management System | Health Management Information System | Web | E-commerce / Telecom. platform | Internet Forums | Email, Recorded Media | Rural Kiosk Machine (RKM) |
| Infrastructure Requirements | Internet, LMS, E-learning Centers | FWT, PDAs, HS, DHS | Solar Power, Internet, Web | Open Source Edubuntu platform, Internet, email, Wikipedia | Internet, Portals, Forum, Web | Internet, Email, Recorded Media | Wireless Network, RKM, ICT-TC |
| Initial Setup Cost | Medium | High | High | Medium | Medium | High | High |

## 6. CONCLUSIONS

It is demand of the day that new educational technologies should be adopted in order to provide a modern education with new teaching methodologies and collaborative learning. ICTs can play important role by introducing new teaching and learning practices thus revolutionizing the educational system. The use of LMSs from higher education to schools is increasing day-by-day, and LMSs are future of educational technologies with a great amount of new possibilities. We present a model for development of e-learning centers in rural areas based upon open-source LMSs with technical overview and features. The study may contribute in promoting education in rural areas by offering latest educational technologies and developing students' interest in e-learning. This will create productivity and contribute to economy thus improving livelihood of rural population. The study may also promote ICTs, e-learning and usage of LMSs. Use of open-source software in proposed model may acquaint learners with potential benefits and motivate learning communities for use of open-source technologies. It may also help in promotion and development of open-source software.

Considering the increasing popularity of e-learning, it is quite worthwhile to offer such facilities in rural areas in order to reduce the digital divide and to support in the process of rural development by contributing in the socio-economic factors.

## 7. ACKNOWLEDGMENTS

The work reported in this paper was supported by the National Natural Science Foundation of China (Grant No. 60903003), the Beijing Natural Science Foundation of China (Grant No. 4112037), and the Research Fund for the Doctoral Program of Higher Education of China (Grant No. 2008000401051).